# RPSE: REIFICATION AS PARADIGM OF SOFTWARE ENGINEERING

**Dr. Viktor Sirotin**

**www.sirotin.eu**

**ABSTRACT**

The paper introduces RPSE, **Reification as a Paradigm of Software Engineering**, and enumerates the most important theoretical and practical problems of the development and application of this paradigm.

## 1. BASIC DEFINITIONS

Before pitching the thesis of this article, it is necessary to agree on the meaning of the main terms.

### 1.1. Software Engineering

We will use the classical definition of software engineering from the IEEE dictionary [1]: “Software engineering is the application of a systematic, disciplined, quantifiable approach to the development, operation, and maintenance of software”.

### 1.2. Paradigm

For the term **paradigm**, we will rely on the classical definition from Thomas Kuhn [2]. According to Kuhn, a paradigm is a set of concepts, universally recognized rules, models, and methods of solving problems in a certain field of science.

The dualism of this concept lies on the one side in the community of experts who accept a certain paradigm. On the other, the acceptance of a certain paradigm means for a specialist to join such a community.

Thomas Kuhn considered in his book scientific paradigms. However, the usefulness of this concept very soon manifested in different areas of technology and social life. Numerous publications about paradigms and their influence began to appear in the specialized and popular literature about the automotive industry, urban planning, the treatment of certain diseases, etc.

Software engineering, and especially its important part programming, were no exception. Now there are many competing programming paradigms. Wikipedia [3] devoted an article to an overview of them; there are as well interesting reviews such as [4]. The paradigms covered focus on only one part of software engineering, namely writing programs in one or another programming language. Each professional software engineer knows that most real software projects cannot be performed with only one of paradigms (for example, functional programming).

The paradigm described in this article, on the contrary, is applicable all areas of software engineering.

Some papers, for example [5], call paradigms different approaches for software project management, e.g., waterfall models or the agile model. From my point of view, these approaches are not acceptable as paradigms in the spirit of the Kuhn's definition due to their relative theoretical simplicity and the absence of a broad useful theoretical basis after many years' existence[1].

### 1.3. Reification

We use the term “**Reification**” in this article as an extension of the classical definition of reification in computer science: “Reification is the process by which an abstract idea about a computer program is turned into an explicit data model or other object created in a programming language” [6].

[1]The paradigm proposed in this article does not yet have its own well-developed theoretical basis. However, ways to build this basis are already visible today.

Consider the process of transition from the abstract (ideal) to the concrete (material) in more detail.

Already in the earliest of the philosophical tracts that have reached us it was customary to oppose the material to the ideal. We can at best feel (or think that we feel) ideal objects, that is, ideas. Indicators of such a sensation of ideal objects can be a change in mood or in the course of thoughts after listening to a piece of music, reading a fragment of a book, etc.

Attempts to formulate our sense of the Ideal as a rule do not lead to success. Even practical ideas such as minor repairs around the house or preparing a new variation of a familiar dish are difficult to formulate at first. Only after thinking about or trying to explain it to another person does the idea acquire more and more clear “outlines”.

Let us now turn from the consideration of the concept of the ideal to the consideration of the material. We can sense and register material objects around us, distinguish their properties qualitatively. The properties of many objects are measurable. We can also objectively identify hierarchies and other structures of material objects. To register or measure an object’s properties it is not even necessary to have an object. It is enough to have its model. Moreover, in many practical situations the model can be used without an object. Models can be discussed with other people. People can negotiate based on models. Models can be standardized (formalized).

In some areas of human activity, the standardization of models has gone so far that physical objects (e.g., a bolt) made from a standardized model (e.g., a technical drawing) by different people or automata are indistinguishable from each other.

Given the relative inaccuracy of the proposed definition, I will divide the world of phenomena of our inner and outer world U into two parts:

$$U = M + I,$$

where the set M consists of phenomena that can be objectively recorded or measured (the material world), and I is everything else.

Whether this formula is applicable to all phenomena of the surrounding world is an open philosophical question. In this article, we will narrow this formula to the world of software engineering.

Each process of creation or update some software system begins with the emergence of some ideas about the future system or its parts in the minds of inventors, customers, or developers. We call these ideas **mental models**.

For simple systems or simple extensions or updates of large systems, the developer immediately writes the code or configures the system based on his or her mental model. However, in most cases people create and use various intermediate models from a simple requirements list to extensive formal models such as UML or BPMN models.

Thus, the process of creating or updating software systems can be represented as a chain of mental and material models:

$$I_1, I_2, ..I_n, M_1, M_2, ..M_m$$

## 2. REIFICATION IN THE OTHER AREAS

It is clear that the above definition is not radically new, and is widely used in the neighboring programming engineering areas of intellectual work, consciously or unconsciously. Consider for example two such areas, mechanical engineering and mathematics.

These two areas have used reification effectively since long ago. They have something that programming engineering can learn.

In mechanical engineering, we see a complete cycle of reification, from the emergence of an idea in the brain of the designer through its details, mapping into a model, and finally material construction.

Reification in mechanical engineering really ends with the creation of material objects rather than formal models.

The situation is different in mathematics[2]. The “final product” of mathematics is a formal model with strictly proven properties.

Mathematics uses a greater number of very abstract models.

Mechanical engineering on the contrary uses relatively fewer abstract models but more specific ones, for example models for CNC machines, which allow production of physical objects.

From this point of view, programming lies in the middle. The final product of programming is program code. In addition, while program code launches on hardware its specific physical objects (electrical signals) are difficult to compare with nuts, gears, and car bodies. On the other hand, program code is close to mathematical formulas; sometimes there is a direct mapping. However, in any real software system you need to take into account many specific aspects of the environment and interaction with users or other systems. This makes program code more specific than mathematical formulas.

Thus, looking at these three neighboring areas, one can speak of an ordering from the point of view of abstractness vs. concreteness of their models:

Abstract <= Concrete

Mathematical formulas <= Program code <= Mechanical engineering models.

### 2.1. What can software engineering learn from neighboring areas in terms of using models?

Let us first consider mathematics.

Over the several thousand years of its development, mathematics has learned to describe the same phenomena of a real or imaginary world from different points of view and in very different terms. The ancient Greeks learned to substitute a purely verbal problem formulation with geometric figures and so to solve practically important problems. Later there was an understanding of the interchangeability of lines on the plane and numbers. Then the concept of an algebraic variable and the reduction of geometric problems to systems of algebraic equations crystallized.

Mathematics has not only learned to describe real or imaginary objects and processes with the help of models of very different mathematical nature. An important achievement of mathematics is the ability to determine the degree of similarity of models from different areas of mathematics and the ability to transform them into each other. Many breakthrough solutions to the most important mathematical problems of recent years are chains of individual proofs, each of which uses specialized apparatus from a special area of mathematics.

In programming, something similar happens now by compiling the source code of the program and by generating code from a DSL (Domain Specific Language) or metadata. However, the culture of working with various model types in software engineering is far behind the mathematical one.

What can software engineering learn in terms of reification from mechanical engineering?

In a mechanical industry, there are chains of coordinated quasi-formal and formal models that end with models, based on which physical objects are manufactured and mounted. For most types of intermediate models are formal methods to verify their validity. Models are the main communication language of specialists with different profiles in the design and manufacture of engineering products.

The situation in IT looks much worse. Only within very large concerns do there exist sets of models and processes. Small firms and IT startups as a rule not only do not have developed formal models and processes, but also do not even suspect their existence. This situation derives from:

- Insufficient effectiveness of existing models and processes
- Insufficient popularity of formal models outside of large concerns
- Disadvantages in the education of developers and especially managers
- Gaps between university education and the real needs of software engineering

---

[2] See interesting notes to this theme in paragraph 7.3 of the book of Alexandre Borovik “Mathematics under the microscope” [7].

## 3. DEFINITION OF RPSE

We have defined all the necessary terms to give the basic definition of the proposed paradigm:

**Software engineering is the reification (materialization of ideas) via the transformation of mental models into code executed on computers**[3].

Within the proposed paradigm:

1. All basic processes of software engineering are concrete variants (implementations) of the process of constructing chains of mental and material models $I_1, I_2,..I_n, M_1, M_2, ..M_m$. The last most specific model in this chain is, as a rule, program code.

2. The essence of software engineering is the construction of such chains.

3. All main issues of optimizing the development, its cost, and quality can be reduced to the optimization of construction of the corresponding chain of models.

The above definition has variations in different areas of software engineering.

Only in a very small number of cases does the developer have at the beginning a full, clear plan of how to solve the problem. In most real projects, the search for a solution and its implementation coexist, evolving in parallel and interacting with each other. The mental models, the code, and the intermediate models (text, images, and formal models such as UML) grow and change together, influencing each other.

Very often, several people work on a problem simultaneously. Each of them has their own mental model and perhaps its intermediate models and code fragments.

Often code in some programming language does not actually exist, the solution reducing to managed masks of configurators or generators.

Finally, computation devices have significantly expanded in recent years. Earlier processor chips were on the boards inside desktops, laptops and server racks. Now there are various chips inside mobile phones, game consoles, surveillance cameras, smart home appliances, etc.

Our RPSE definition is applicable to all the areas listed above.

## 4. CONTOURS OF RPSE

Is it possible to outline the RPSE paradigm's contours more precisely?

The next step is an attempt to outline the main categories of phenomena that it affects. Remembering Kuhn's definition, we need to list the types of models that introduce and use the paradigm.

Models of RPSE can be divided into three main categories:

- Mental models

- Code in programming languages or its equivalents as a model of executable code.

- Intermediate models

The least studied in this triad are the mental models. What exactly do they mean?

**Mental model** is a term for designating ideas that exist in the minds of customers, programmers and other participants in the software engineering process. The existence of such models is indisputable and can be registered by the programmer him- or herself. At the modern level of technology development, these models cannot be well measured by instruments.

One indirect method to extract and measure such models that works well is an interview with the idea holder. Obviously, the process of interviewing dramatically affects the mental model itself. Each of us has probably experienced the situation when an attempt to formulate a problem for a colleague led to “enlightenment”, and often the solution of the problem.

[3] Note that although the main and more detailed definitions of the paradigm speak about the construction of chains of models, the paradigm is called reification rather than modeling. This is because the essence of these chains is that they become less abstract and increasingly more concrete.

Based on correctly formulated questions, an interview allows objectively construction of models of varying complexity. The most common such models are[4]:

- Sets or lists with binary, enumeration, numeric, string and other elements,
- Graphs
- Formal models for behavior definition (e.g. finite-state machines)
- Semiformal behavior models

## 5. WHY DOES SOFTWARE ENGINEERING NEED A CROSSCUTTING PARADIGM?

A crosscutting paradigm like RPSE opens opportunities for participants in the software engineering process:

- The opportunity for all participants in the process to use the same terminology.
- The ability to systematically build an end-to-end process for creating new software, based on well-defined models.
- The ability to objectively evaluate intermediate results of the software engineering process and to manage it.

## 6. THE MAIN TASKS FOR THE DEVELOPMENT OF THE PARADIGM

### 6.1. Theoretical problems

As noted in Kuhn’s book [2], in most cases scientists are engaged in the solution of problems in an environment of potential solutions from existing paradigms, and rarely undertake ones where it is not very clear how to approach them. Nevertheless, these are our tasks, mainly:

1. Constructive definition of the concept of the mental model.
2. Finding constructive criteria for evaluating the degree of abstraction (ideality) vs. specificity (concreteness) of models.
3. Finding criteria for selecting candidate intermediate and additional models.
4. Selection and development of criteria and methods for comparing models of different types, including their direct and reverse tracing.
5. Development of methods for automated and automatic transformation of models.

### 6.2. Practical tasks

Along with the theoretical tasks for the development of the described paradigm, for the introduction of RPSE it is necessary to solve at least the following practical problems:

---

[4] These models are a reflection of mental models. The degree of their closeness to real mental models should be handled by psychology or pedagogic theory. Unfortunately, the author does not know of serious work in this area. This does not mean that such work does not exist.

1. Creating tools for:
    a. Extraction of mental models
    b. Automated and automatic transformation of mental models into intermediate models or code in programming languages.
    c. Tracing and monitoring the essential changes in the model content by model transformation.
2. Writing of necessary technical, educational literature and teaching materials.
3. Organization of professional permanent forums and regular RPSE-conferences.

## 7. CONCLUSION

This paper attempts to define the RPSE paradigm of software engineering as the reification (materialization of ideas). The word “define” is not used here by chance. In fact, participants in IT projects have long been engaged in the creation, transformation, and use of models (most probably, subconsciously).

In the strict sense of Kuhn's definition, the described approach does not have the right to be called a paradigm because it does not have a vast community of people supporting it or a developed system of interrelated models. However, there is hope that from solving the tasks listed in paragraph 6 of this paper this will happen.